\documentclass[sigconf]{acmart}

\copyrightyear{2026}
\acmYear{2026}
\setcopyright{cc}
\setcctype{by}
\acmConference[DaMoN '26]{22nd International Workshop on Data Management on New Hardware}{May 31-June 05, 2026}{Bengaluru, India}
\acmBooktitle{22nd International Workshop on Data Management on New Hardware (DaMoN '26), May 31-June 05, 2026, Bengaluru, India}
\acmDOI{10.1145/3789237.3809130}
\acmISBN{979-8-4007-2455-8/2026/05}

\newcommand{\labeltitle}[1]{\emph{#1}.}

\begin{document}

\title{Should I Hide My Duck in the Lake?}

\author{Jonas Dann}
\email{jonas.dann@inf.ethz.ch}
\affiliation{%
  \institution{ETH Zürich}
  \city{Zürich}
  \country{Switzerland}
}

\author{Gustavo Alonso}
\email{alonso@inf.ethz.ch}
\affiliation{%
  \institution{ETH Zürich}
  \city{Zürich}
  \country{Switzerland}
}

\begin{abstract}

Data lakes spend a significant fraction of query execution time on fetching and scanning data from remote, disaggregated storage.
Data decoding alone accounts for $46\%$ of runtime when running TPC-H directly on Parquet files. 
To address this bottleneck, we propose a vision for a data processing SmartNIC for the cloud that sits on the network datapath of compute nodes to offload decoding and pushed-down operators, effectively hiding the cost of parsing raw files. 
Our experimental estimations with DuckDB suggest that by operating directly on pre-filtered data, as delivered by a SmartNIC, we can significantly increase query processing performance and can still match query throughput of traditional setups with smaller, less expensive CPUs.

\end{abstract}



\maketitle

\section{Introduction}
\begin{figure}[bt]
	\centering
	\includegraphics[width=\linewidth]{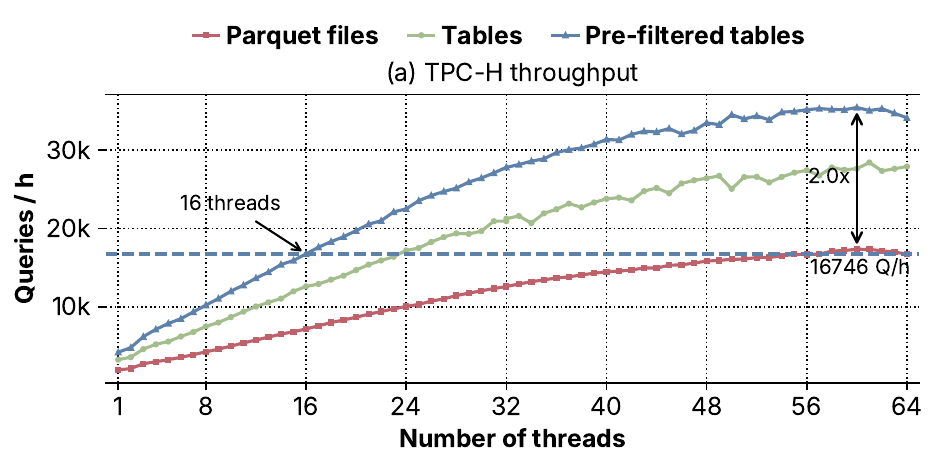}
	\caption{DuckDB TPC-H (scale factor 30, 4 streams) throughput benchmark on Parquet-resident data, pre-loaded tables, and pre-filtered tables.}
	\label{fig:throughput}
\end{figure}

Cloud-native database systems have addressed key limitations of traditional architectures---elasticity, cost efficiency, and resource pooling---by disaggregating compute and storage.
However, disaggregated architectures introduce their own challenges around data movement. 
Two recent studies of production data warehouse workloads show that table scanning and filtering alone account for roughly half of total query runtime~\cite{journals/pvldb/SzlangBCDFHOOM25, journals/pvldb/RenenHPVDNLSKK24}, rising to over 80\% for read-only queries~\cite{journals/pvldb/RenenL23}.
In data lakes, which process queries directly on raw input files circumventing a previous data ingestion step, this bottleneck is further amplified by ad-hoc per-query decoding of encoded and compressed formats like Parquet~\cite{parquet} and parsing of raw text formats like CSV and JSON.
As storage and network bandwidths have increased by orders of magnitude over the last decade, the bottleneck in these cloud-native database systems increasingly shifts to the CPU~\cite{journals/pacmmod/KuschewskiGNL24, journals/pvldb/ZengHSPMZ23, conf/nsdi/VuppalapatiMATM20, conf/cidr/HuB0025}.
Figure~\ref{fig:throughput} shows that data decoding and filtering severely limit query throughput in DuckDB \cite{conf/sigmod/RaasveldtM19} and performance scaling with many threads becomes increasingly inefficient.
Notably, for TPC-H, 16 threads on pre-filtered data outperform the full 64 cores of the CPU of our evaluation setup on Parquet-encoded data.
Utilizing all cores of the CPU, DuckDB achieves twice the query throughput on pre-filtered data compared to Parquet-encoded data.
Pre-filtered data refers to replacing pushed-down filters in scan operators and filter operators sitting directly above scan operators with a scan on a pre-materialized view (cf. Figure~\ref{fig:view_rewriter}).

\begin{figure*}[bt]
	\centering
	\includegraphics[width=\textwidth]{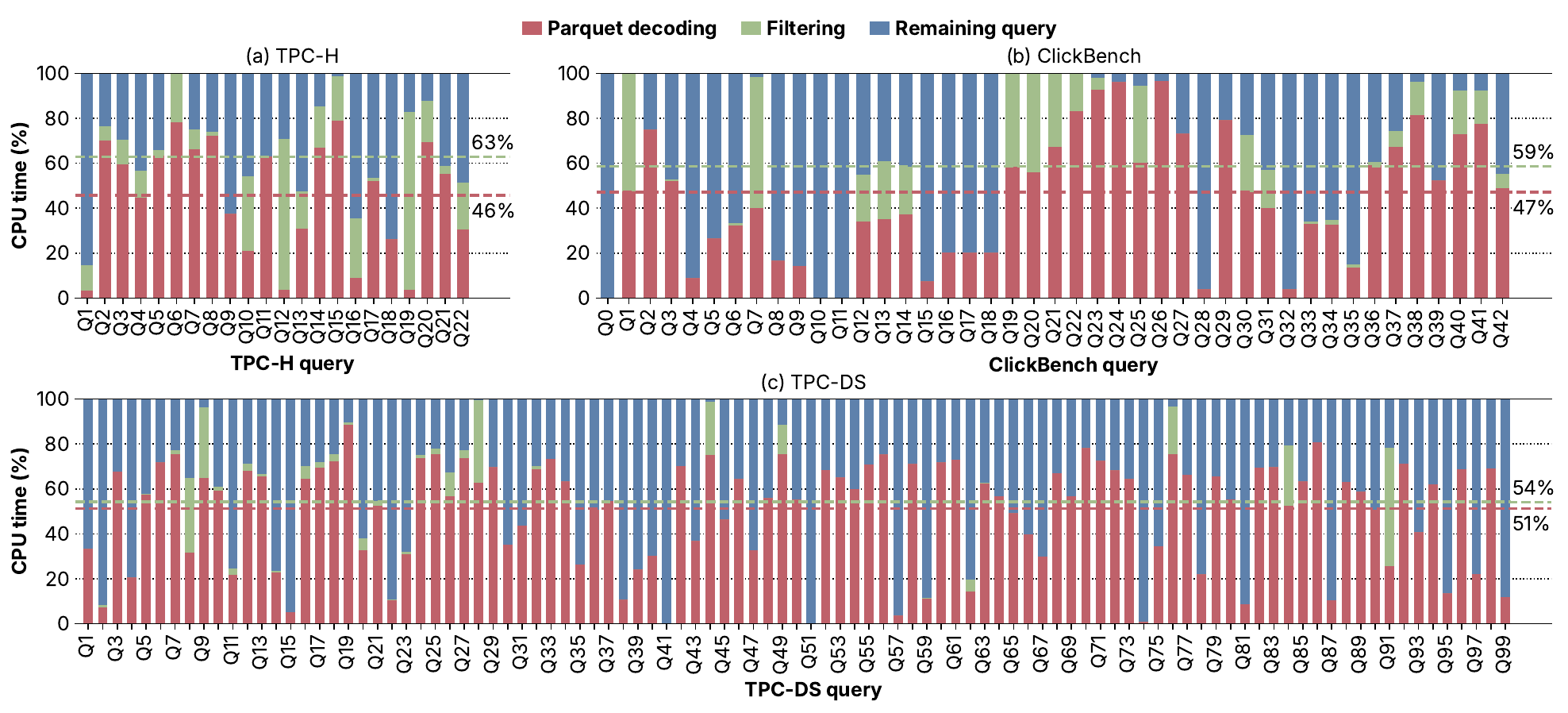}
	\caption{Per-query breakdown of Parquet decoding, filtering, and the remaining query runtime for the benchmarks TPC-H (scale factor 30), ClickBench, and TPC-DS (scale factor 30).}
	\label{fig:stacked}
\end{figure*}

Cloud-native systems have adopted many different approaches to mitigate this high cost of scanning and filtering data.
To reduce the pressure of scanning in the compute layer, operator pushdown into the storage layer \cite{conf/icde/YuYWGSAS20, journals/vldb/YangYSAS24} and intermediate layers that partially evaluate queries \cite{conf/icde/YuYWGSAS20, journals/debu/CaiGGNPPP18, conf/sigmod/ArmenatzoglouBB22} have been explored successfully.
Notable commercial systems that employ operator pushdown are Google BigQuery, which pushes filter and projection evaluation into its Capacitor storage layer~\cite{journals/pvldb/0001GLRSTVADMPS20}, and Oracle's Exadata appliance, which offloads predicate filtering, column projection, and join Bloom-filters to storage~\cite{exadata-smartscan}, in both cases significantly reducing the volume of data that must cross the storage-compute boundary.
On the storage side, the decoding overhead of raw data files has motivated new formats using nested, lightweight compression schemes, SIMD-friendly data organization \cite{journals/pacmmod/KuschewskiSAL23, journals/pvldb/AfroozehB25, journals/pvldb/GienieczkoKNLG25}, and aggressive file pruning \cite{conf/sigmod/ZimmererDKWOK25}.
On the hardware side, dedicated accelerator prototypes have demonstrated the potential of accelerated Parquet-to-Arrow decoding \cite{conf/icfpt/PeltenburgLH0AH20} and line-rate JSON parsing \cite{conf/fpt/PeltenburgHBMA21, conf/damon/DannW0FF22}.
However, these accelerator approaches remain isolated efforts without a unified system-level integration into cloud-native database systems.
SmartNICs, in particular, offer a unique opportunity for data lakes to offload data processing from the CPU into the network datapath \cite{journals/access/KfouryCMAGC24}.
Cloud providers already deploy SmartNICs at massive scale.
For example, Microsoft Azure has been deploying FPGA-based SmartNICs for network virtualization in all new servers for over a decade \cite{conf/nsdi/FirestonePMCDAA18}.

In the remainder of this paper, we first discuss a detailed study of the cost of data decoding and filtering across the TPC-H, ClickBench, and TPC-DS benchmarks.
We show experiments collected with DuckDB that analyze the per-query overhead of decoding and filtering, and the effect of different input file configurations on query performance.
We identify the potential to significantly reduce these overheads and largely hide the cost of processing queries on raw input files in the network datapath by offloading data decoding and pushing down operators to a SmartNIC.
Finally, we propose a vision for a data processing SmartNIC for cloud-native database systems.
Our approach enables substantial reductions in provisioned compute resources and improves overall resource efficiency in cloud-scale data lake deployments.
We further identify and discuss the open research challenges that we will address in future work to realize this vision.

\section{Query Performance on Raw Data}
\begin{figure}[bt]
	\centering
	\includegraphics[width=\linewidth]{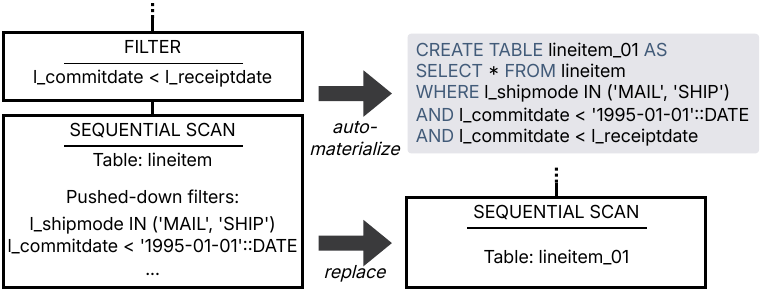}
	\caption{DuckDB scan rewriter optimizer extension.}
	\label{fig:view_rewriter}
\end{figure}

All experiments shown in this paper are collected with DuckDB 1.4.4 with varying input configurations.
To emulate the pre-filtered data that a SmartNIC would pass to the database system, we built a DuckDB extension that rewrites query plans with a post-optimizer hook and replaces table scan operators with pushed down filters and filter operators sitting directly above a scan with scans of pre-materialized tables (cf. Figure~\ref{fig:view_rewriter}).
This ensures identical query plans across all measurements so they stay comparable.
When the extension is active, all queries of the currently used benchmark are executed once as a warmup run to auto-materialize all pre-filtered views with a second, out-of-band connection to the database.
We also add replacement rules based on the filter fingerprint to an internal ruleset that is applied whenever the same filter is encountered again.
We run all benchmarks on a dual-socket server equipped with two AMD EPYC 7V13 CPUs ($64$ cores, no hyper-threading) and 512\,GB of memory.
We pin DuckDB's threads and memory allocation to socket $0$.
Parquet, CSV, and JSON files are stored in a tmpfs to isolate file decoding and parsing from SSD latency and bandwidth.
DuckDB and our benchmark code\footnote{\href{https://github.com/celeris-labs/duckdb-bench-damon2026}{https://github.com/celeris-labs/duckdb-bench-damon2026}} are compiled with GCC~11.4.0.
Each data point reports the median of five runs.

Figure \ref{fig:stacked} breaks down per-query performance of the TPC-H, ClickBench, and TPC-DS benchmarks into three components: Parquet decoding, filtering, and remaining query runtime.
On average, decoding accounts for roughly half of the runtime per query across all benchmarks.
Filtering contributes an additional 17\% and 12\% for TPC-H and ClickBench, respectively.
For scan-heavy queries with very high selectivity such as TPC-H Q6 and Q15 or ClickBench Q7 and Q21, these two components dominate the entire runtime. 
For aggregation- and join-heavy queries like TPC-H Q1 and Q16 or ClickBench Q11 and Q32, scanning makes less of a difference to the runtime.
The TPC-DS queries also show roughly half of the runtime being used for Parquet decoding.
However, even though there are many pushed-down filters, they make up significantly less of the query runtime compared to TPC-H and ClickBench.
This is because pushed-down filters are usually applied to the smaller tables in the data set and, for larger tables, tuples are only filtered out while joining.
We expect that Bloom filter push-down will make a big difference for these queries.
These distributions remain consistent across scale factors and thread counts for TPC-H and TPC-DS.

\begin{figure}[bt]
	\centering
	\includegraphics[width=\linewidth]{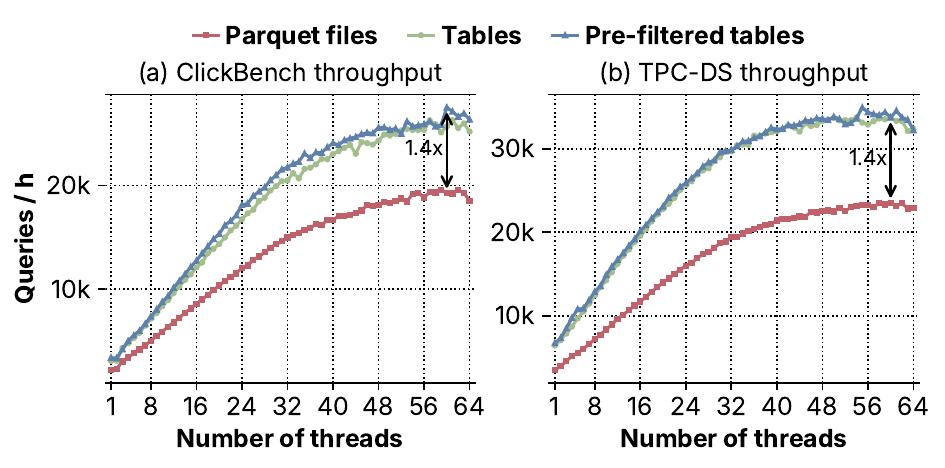}
	\caption{DuckDB ClickBench (4 streams) and TPC-DS (scale factor 30, 4 streams) throughput benchmark for Parquet-resident data, pre-loaded tables, and pre-filtered tables.}
	\label{fig:throughput-other}
\end{figure}
Based on these observations, we would expect ClickBench to benefit significantly from pre-filtered tables in a throughput benchmark.
Figure~\ref{fig:throughput-other} shows the throughput benchmarks for ClickBench and TPC-DS.
Similarly to TPC-H, Parquet decoding slows down query throughput significantly.
However, pre-filtered tables do not improve performance significantly for either benchmark.
For TPC-DS, we already observed this in the per-query breakdown.
However, for ClickBench, this is because queries that do not benefit from a pre-filtered table, especially Q28 and Q32, dominate over half of the total runtime of the benchmark.

\begin{figure}[bt]
	\centering
	\includegraphics[width=\linewidth]{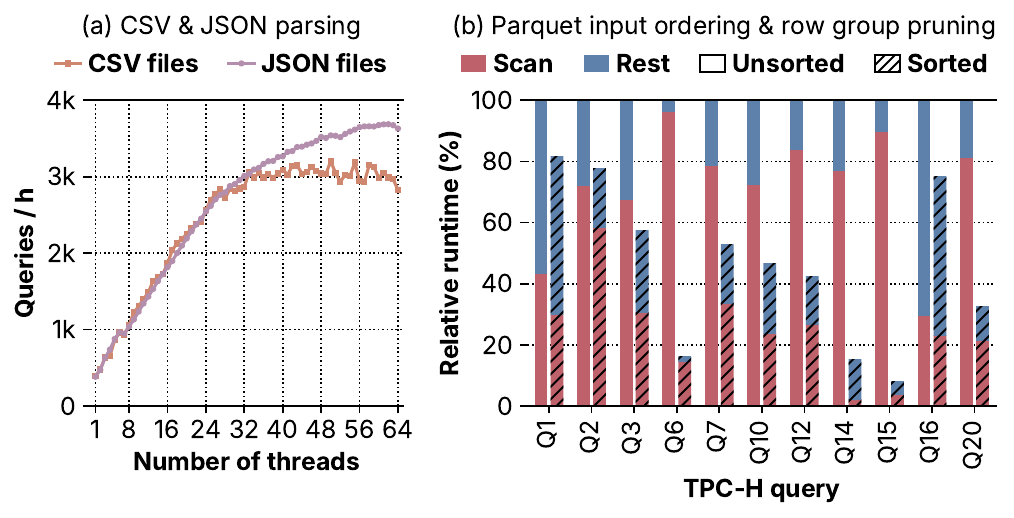}
	\caption{TPC-H CSV and JSON throughput (scale factor 10) \& effects of Parquet input ordering (scale factors 30).}
	\label{fig:csv}
\end{figure}

The cost of scanning becomes even more pronounced for text-based formats.
Figure~\ref{fig:csv}(a) shows query throughput when executing TPC-H directly on CSV and JSON files.
Both data formats as sources perform nearly identical up to $32$ threads, after which JSON parsing scales slightly better, while CSV parsing plateaus. 
Overall, throughput for both formats is dramatically lower than even for Parquet, where we see $14-16\times$ higher throughput. 
While Parquet decoding already implies a significant overhead, text-based formats---lacking columnar organization, binary encoding, and predicate-relevant metadata---are far more costly to parse.
Figure~\ref{fig:csv}(b) compares per-query runtime on Parquet files with a random permutation (unsorted) and sorted data\footnote{Sorting \texttt{lineitem} on \texttt{l\_shipdate} and \texttt{orders} on \texttt{o\_orderdate}} for queries where ordering made more that 10\% difference in runtime (we exclude the rest of the queries to make the plot more readable).
Sorting allows for row group pruning in Parquet based on zone maps.
For scan-heavy queries that use the sorted columns in filters, this results in a significant performance improvement (e.g., Q6, Q14, and Q15).

\labeltitle{Discussion}
Across all three benchmarks, three overarching insights emerge. 
First, the cost of data decoding and filtering dominates query execution: Parquet decoding alone takes around half of CPU runtime per query. 
This cost is fundamental to data lake systems and independent of the particular benchmark. 
Second, the actionable fraction of this cost varies substantially with workload characteristics. 
TPC-H benefits the most from pre-filtering because its pushed-down predicates have high absolute selectivity, yielding the full $2\times$ throughput improvement. 
TPC-DS pushes filters mostly into smaller dimension tables which does not significantly reduce the amount of data passed on at the scan level.
Finally, format- and layout-level characteristics (text vs. binary, unsorted vs. sorted) produce order-of-magnitude differences in scan cost, suggesting that any solution targeting the data decoding bottleneck must handle a heterogeneous mix of file formats. 
Together, these observations indicate that decoding and filter push-down are both costly enough to warrant offloading and structurally lend themselves to it.

\section{Hiding Queries in the Datapath}
\begin{figure}[bt]
	\centering
	\includegraphics[width=.8\linewidth]{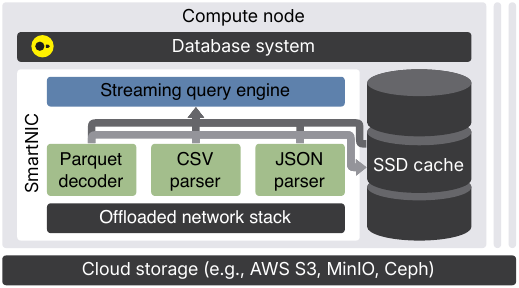}
	\caption{Data processing SmartNIC architecture.}
	\label{fig:idea}
\end{figure}

Figure \ref{fig:idea} illustrates our vision for a data processing SmartNIC architecture for next-generation cloud-native database systems based on the SCENIC SmartNIC \cite{journals/corr/abs-2604-15128}.
Positioned in the compute nodes on the datapath between cloud object storage and a host database system, the SmartNIC decodes raw input files at line-rate, eliminating the decoding bottleneck before data ever reaches the CPU.
The decoded data streams are then routed through an accelerated on-NIC query engine for pushed-down operator evaluation and simultaneously placed in an SSD-based table cache for efficient data reuse.
The accelerated query engine filters the data streams and optionally applies streaming operators, for example, arithmetic projection expressions, Bloom filters for probe-side filtering in joins, pre-aggregation of results, or join operators where one relation fits entirely in on-chip memory.
On top of these, more sophisticated operators may explore deeper integration of the host database system and SmartNIC, optimizing scheduling of data and in-memory data structures for CPUs and query processing on GPUs.
Realizing this vision presents three primary open challenges of SmartNIC hardware-software co-design.

\labeltitle{Line-rate data decoding}
The first challenge is decoding various storage formats at 100G and beyond network line-rate \cite{conf/sosp/RamhorstKHDLA25, journals/corr/abs-2507-20412}. 
Formats like Parquet layer multiple encoding schemes whose interactions create complex decoding pipelines.
Text-based formats pose challenges with quoting and escaping, numbers encoded as strings, and transposing row-based records for columnar database systems.
Decoders for different formats should share as many resources as possible to maximize performance.
Furthermore, accessing object storage directly from an accelerator remains an open challenge.

\labeltitle{From decoded data to query processing}
The second major challenge is the semantic gap between network stack offloading and analytical query processing.
Prior work on accelerated query processing \cite{journals/pvldb/WoodsIA14, conf/cidr/KorolijaKKTMA22, conf/fccm/OwaidaSKA17} has demonstrated operator offloading, but targets slow, locally-attached storage over PCIe with rigid schemas and queries.
Bridging this gap requires an accelerated, streaming query execution engine that evaluates pushed-down operator pipelines with runtime schema and query flexibility and collaboratively manages memory with the host database system to enable zero-copy handoff of intermediate results.
Additional open questions include operator-level parallelism \cite{journals/pvldb/GiouroukisNPZM25} and hybrid NIC-GPU cross-accelerator execution~\cite{journals/csur/DannRF23} which offers exciting opportunities~\cite{journals/pvldb/KabicWDA25}, especially for vector search and AI workloads.

\labeltitle{SSD table cache}
The third challenge is managing a local table cache on direct-attached SSDs (i.e., without CPU involvement in the data movement) to avoid redundant network fetches for frequently accessed data.
SSDs may either be accessed through PCIe or directly plugged into the accelerator board.
Studies show that a lot of queries are 1-on-1 repetitions of previously seen queries \cite{journals/pvldb/RenenHPVDNLSKK24}.
Two recent FPGA-based prototyping platforms have demonstrated accessing SSDs through PCIe without CPU involvement~\cite{journals/corr/abs-2604-15128, journals/corr/abs-2503-09318}.
A complementary line of in-storage processing work has pushed query execution itself into the SSD---from operator-level offloading \cite{journals/pvldb/JoBYKCLJ16} to running a full SQL engine inside the device \cite{journals/pvldb/ParkCOL21}---but these efforts assume a host-attached SSD as the sole data source rather than a caching layer behind a network-facing accelerator.
In our setting, it is unclear how to manage metadata and orchestrate SSD and network traffic as two data sources for the streaming query execution engine.
Lastly, there are exciting opportunities to explore different storage formats for the SSD cache.

\labeltitle{Concurrency \& multi-tenancy}
The line-rate decoders, streaming query engine, and SSD cache all contend for the same on-NIC resources (i.e., compute, on-chip caches, memory bandwidth, and PCIe and SSD I/O). 
Without explicit multi-query scheduling, bandwidth partitioning, and QoS, a scan-heavy query can easily starve concurrent RDMA traffic or stall a parallel decode pipeline, degrading the benefits that motivate datapath offloading. 
This cross-cutting nature distinguishes our setting from prior accelerated query engines, where decode, operator evaluation, and storage typically lived on separate devices.
A full multi-tenant deployment, in which independent database instances share a single SmartNIC with isolation and fairness guarantees, raises further questions around context switching, partial reconfiguration, and per-tenant cache accounting that we consider out of scope for an initial prototype. 
We plan to first demonstrate the single-tenant case end-to-end and explore a multi-tenant extension in future work.

\section{Conclusion}
Data lakes are bottlenecked by the CPU cost of decoding and filtering raw input files.
Parquet decoding consumes around half of per-query runtime across TPC-H, ClickBench, and TPC-DS. 
Our experimental evaluation with DuckDB shows that operating on pre-filtered data---the kind of stream a SmartNIC could deliver---yields up to $2\times$ higher TPC-H throughput and lets 16 CPU cores match the performance of 64 cores on Parquet-resident data. 
We propose a vision for a data processing SmartNIC that realizes this opportunity by combining line-rate decoders, a streaming on-NIC query engine, and a local SSD table cache.
To realize this vision, we identify the open challenges of heterogeneous decoding, bridging the semantic gap to analytical query processing, cache management, and concurrent, multi-tenant resource sharing. 
We will pursuing these directions in an end-to-end prototype on SCENIC as a path toward more efficient cloud-native data lake systems.

\section*{Acknowledgements}
We would like to thank AMD for the donation of the Heterogeneous Accelerated Compute Cluster (HACC) at ETHZ which was used to obtain the experimental results presented in this paper and Geert Roks for the support of the cluster. 
This work was funded in part through an unrestricted grant from AMD.

\bibliographystyle{ACM-Reference-Format}
\bibliography{main.bib}

\end{document}